\documentclass[runningheads]{llncs}
\usepackage{graphicx}
\usepackage{multirow}
\usepackage{amsmath, amssymb}
\usepackage{makecell}
\usepackage{booktabs}
\usepackage{hyperref}
\usepackage{color}

\usepackage[table]{xcolor}
\usepackage[misc]{ifsym}

\def\eg{\emph{e.g.}}

\begin{document}
\title{Arbitrary Reduction of MRI Inter-slice Spacing Using Hierarchical Feature Conditional Diffusion}
\titlerunning{HiFi-Diff: Hierarchical Feature Conditional Diffusion}
% \thanks{Supported by organization x.}}

\author{
Xin Wang\inst{1}* \and 
Zhenrong Shen\inst{1}* \and 
Zhiyun Song\inst{1} \and 
Sheng Wang\inst{1} \and 
Mengjun Liu\inst{1} \and 
Lichi Zhang\inst{1} \and 
Kai Xuan\inst{2} \and 
Qian Wang\inst{3}$^{(\textrm{\Letter})}$
}
% \author{Anonymous}

% \authorrunning{Anonymous}
\authorrunning{X. Wang et al.}

% \institute{Anonymous Organization}
\institute{
School of Biomedical Engineering, Shanghai Jiao Tong University, Shanghai, China \and
School of Artificial Intelligence, Nanjing University of Information Science and Technology, Nanjing, China \and 
School of Biomedical Engineering, ShanghaiTech University, Shanghai, China
\email{qianwang@shanghaitech.edu.cn}
}
% % \email{lncs@springer.com}\\
\maketitle              % typeset the header of the contribution
\renewcommand{\thefootnote}{}
\footnotetext{* Contributed equally to this work.}
% \footnote{* Contributed equally to this work.}

\begin{abstract}
Magnetic resonance (MR) images collected in 2D scanning protocols typically have large inter-slice spacing, resulting in high in-plane resolution but reduced through-plane resolution.
Super-resolution techniques can reduce the inter-slice spacing of 2D scanned MR images, facilitating the downstream visual experience and computer-aided diagnosis.
However, most existing super-resolution methods are trained at a fixed scaling ratio, which is inconvenient in clinical settings where MR scanning may have varying inter-slice spacings.
% Inspired by recent progress in denoising diffusion models in image synthesis, we propose \textit{\textbf{Hi}erarchical \textbf{F}eature Cond\textbf{i}tional \textbf{Diff}usion (HiFi-Diff)} for arbitrary reduction of MR inter-slice spacing.
% Given two adjacent MR slices, HiFi-Diff can iteratively convert a Gaussian noise map into any in-between MR slice.
To solve this issue, we propose \textit{\textbf{Hi}erarchical \textbf{F}eature Cond\textbf{i}tional \textbf{Diff}usion (HiFi-Diff)} for arbitrary reduction of MR inter-slice spacing.
% Given two adjacent MR slices, HiFi-Diff can iteratively convert a Gaussian noise map into any in-between MR slice.
% Specifically, the conditional features are hierarchically extracted from two adjacent MR slices for fine-grained conditioning.
% The relative spatial position of the generated slice is continuously represented for handling different scaling ratios.
Given two adjacent MR slices and the relative positional offset, HiFi-Diff can iteratively convert a Gaussian noise map into any desired in-between MR slice.
Furthermore, to enable fine-grained conditioning, the Hierarchical Feature Extraction (HiFE) module is proposed to hierarchically extract conditional features and conduct element-wise modulation.
% Compared with other arbitrary-scale super-resolution methods, HiFi-Diff produces images of higher quality and clearer structure.
Our experimental results on the publicly available HCP-1200 dataset demonstrate the high-fidelity super-resolution capability of HiFi-Diff and its efficacy in enhancing downstream segmentation performance.

\end{abstract}

\section{Introduction}
Magnetic resonance imaging (MRI) is essential for analyzing and diagnosing various diseases, owing to its non-invasive property and superior contrast for soft tissues. In clinical practice, 2D scanning protocols are commonly employed for MR image acquisition due to limitations in scanning time and signal-to-noise ratio.
% 2D scanning protocols are widely used in MR image acquisition due to the limitation of scanning time and the signal-to-noise ratio in clinical practice.
Typically, such scanning protocols produce MR volumes with small intra-slice spacing but much larger inter-slice spacing, which poses a great challenge for many volumetric image processing toolkits~\cite{freesurfer,FSL} that require near-isotropic voxel spacing of the input images. 
% However, most MR image processing toolkits require near-isotropic voxel spacing on the input images.
Therefore, it is necessary to resample the acquired volumes to align the inter-slice spacing with the intra-slice spacing.

Interpolation methods are widely used to reduce the inter-slice spacing of 2D scanned MR volumes. 
However, these methods simply calculate missing voxels as a weighted average of the adjacent ones, leading to inevitable blurred results.
For better performance, many deep-learning-based super-resolution (SR) studies have been investigated~\cite{SRCNN,DeepResolve,LIIF,ArSSR,MetaSR}.
In this paper, we term large slice spacing as \textit{low resolution} (LR) and small slice spacing as \textit{high resolution} (HR).
DeepResolve~\cite{DeepResolve} adopts a 3D convolutional network to obtain HR images from LR ones, but it only considers reducing inter-slice spacing at a fixed ratio.
Training such an SR model for each scaling ratio is impractical, as it requires significant time and computational resources.

To tackle this issue, local implicit image function (LIIF)~\cite{LIIF} and MetaSR~\cite{MetaSR} are proposed to perform arbitrary-scale SR for natural images.
To achieve arbitrary-scale SR of MR images, ArSSR~\cite{ArSSR} extends LIIF to 3D volumes and utilizes a continuous implicit voxel function for reconstructing HR images at different ratios.
% However, ArSSR tends to generate blurred results compared with the SR models trained on fixed ratios.
% However, like the PSNR-oriented method above, ArSSR also suffers from severe over-smoothing while achieving high quantitative results.
Although ArSSR produces competitive quantitative results, it still suffers from image over-smoothing.
To solve the aforementioned problem, adversarial learning~\cite{goodfellow2020generative} is usually introduced to synthesize more image details. However, such a training scheme often leads to training instability and is prone to generate artifacts~\cite{DBLP,dhariwal2021diffusion}.

\begin{figure}[t]
    \centering    
    \includegraphics[width=\textwidth]{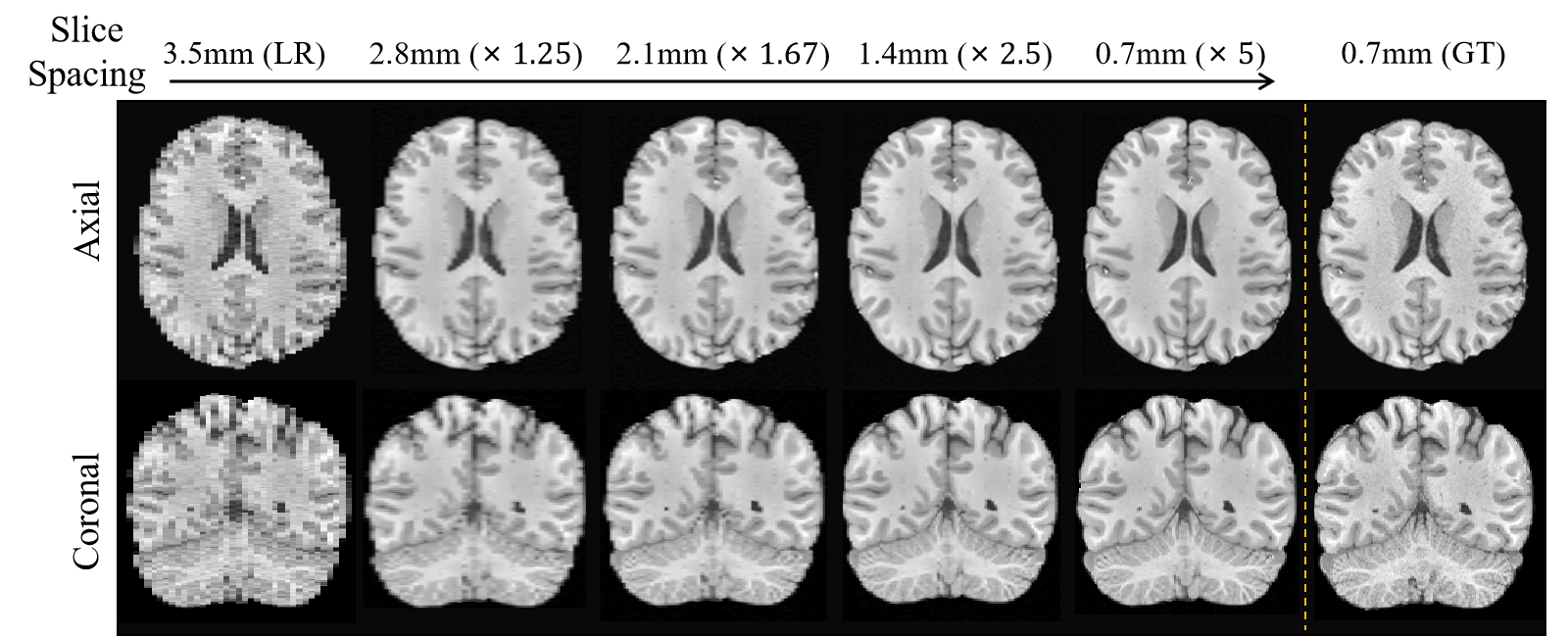}
    \caption{A test case of applying HiFi-Diff to arbitrary reduction of MR inter-slice (sagittal) spacing. The visual quality of the resulting images in the axial and coronal views is gradually enhanced with increasing SR factors.}
    \label{fig:example}
\end{figure}

Recently, diffusion models~\cite{sohl2015deep,ho2020denoising} have achieved wide success in image synthesis tasks, outperforming other deep generative models in terms of visual fidelity and training stability.
Typical denoising diffusion models (\eg, DDPM~\cite{ho2020denoising}) use a series of denoising operations to iteratively generate samples from a prior distribution (\eg, Gaussian) to a desired data distribution.
% first define a forward process where the image is gradually corrupted by various levels of Gaussian noise. 
% An autoencoder is then trained to progressively denoise the unstructured noise map into an image through a Markov chain.
% Based on such a forward-reverse diffusion process, a desired data distribution can be converted from a prior distribution (\eg, Gaussian) via iterative diffusion steps.
Although there exist several works that apply diffusion models for MR image reconstruction~\cite{cui2022self} or denoising~\cite{chung2022mr}, the application of diffusion models to achieve arbitrary-scale MR image super-resolution has not been studied yet.

In this paper, by leveraging the powerful ability of the diffusion models, we propose \textit{\textbf{Hi}erarchical \textbf{F}eature Cond\textbf{i}tional \textbf{Diff}usion (HiFi-Diff)}, which allows arbitrary reduction of inter-slice spacing for 2D scanned MR images, as shown in Fig.~\ref{fig:example}.
Conditioned on two adjacent LR slices, HiFi-Diff can generate any in-between MR slices.
To handle different ratios of inter-slice spacing, we construct continuous representations for the spatial positions between two adjacent LR slices by providing relative positional offsets as additional conditions.
Inspired by the core idea of FPN~\cite{lin2017feature}, we propose the Hierarchical Feature Extraction (HiFE) module, which applies different-scale feature maps as conditions to perform element-wise feature modulation in each layer.
The experimental results demonstrate that HiFi-Diff produces MR slices of excellent quality and effectively enhances downstream image segmentation tasks.
% The contributions of this paper are two-fold. First, we propose a novel HiFi-Diff network for generating high-fidelity slices of MR images. To the best of our knowledge, our proposed HiFi-Diff is the first diffusion model for arbitrary-scale SR of MR images. Second, i

In summary, the main contributions of this paper include:
(1) To the best of our knowledge, HiFi-Diff is the first diffusion model for arbitrary-scale SR of MR images.
(2) We propose the HiFE module to hierarchically extract conditional features for fine-grained conditioning on MR slice generation.

\section{Method}
\begin{figure}[t]
    \centering
    \includegraphics[width=0.9\textwidth]{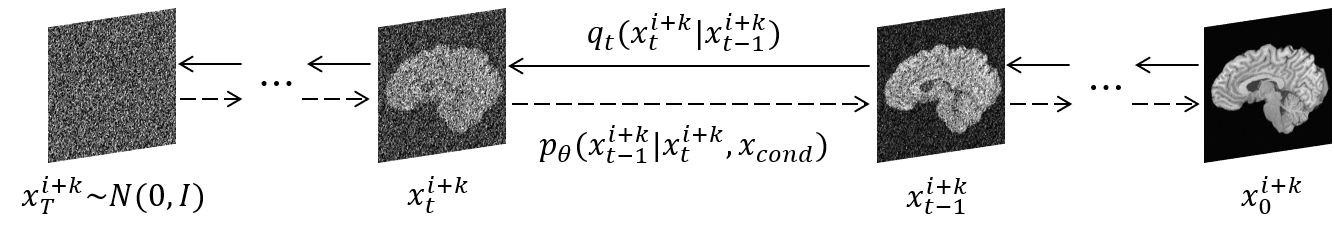}
    \caption{Conditional diffusion process of HiFi-Diff, where $q_t$ and $p_{\theta}$ denote single-step transitions in forward and reverse processes, respectively.
    % Conditional diffusion process of HiFi-Diff, where $q_t$ denotes the forward process and $p_{\theta}$ denotes the reverse diffusion.
    }
    \label{fig:diffusion}
\end{figure}

% Given a pair of adjacent MR slices, HiFi-Diff can generate any in-between slice via iterative diffusion steps.
We discuss the conditional diffusion process and the network architecture of HiFi-Diff in Section~\ref{sec:diffusion} and Section~\ref{sec:network}, respectively.
% The overall framework of the proposed HiFi-Diff is demonstrated in Fig.~\ref{fig:overview}(a), which consists of a \textbf{Hi}erarchical \textbf{F}eature \textbf{E}xtraction (HiFE) module and a main branch for MR slice generation.
% For an LR volume, two adjacent MR slices, along with the offset which represents the relative spatial position between the target slice and the conditional ones, are processed by HiFE module to construct an hourglass-like feature pyramid.
% Conditioned on such a feature hierarchy, the main branch generates the target MR slice via iterative diffusion steps. 

\subsection{Conditional Diffusion for Arbitrary-scale Super-resolution}
\label{sec:diffusion}
Let $x_{0}^{i}\in\mathbb{R}^{H\times W}$ denote a sample from 2D MR slice distribution, where the subscript $0$ refers to the initial timestep and the superscript $i$ refers to the slice index.
For arbitrary-scale SR, we aim to learn continuous representations for the spatial positions between any two adjacent MR slices in an LR volume.
Specifically, we define the generated slice between $x_{0}^{i}$ and $x_{0}^{i+1}$ as $x_{0}^{i+k}$, where $k\in[0,1]$ is a non-integral offset denoting its relative distance to $x_{0}^{i}$.
% The proposed HiFi-Diff is also capable to reconstruct the conditional slices when $k$ is set to 0 or 1.

Similar to DDPM~\cite{ho2020denoising}, HiFi-Diff learns a Markov chain process to convert the Gaussian distribution into the target data distribution, as demonstrated in Fig.~\ref{fig:diffusion}.
The forward diffusion process gradually adds Gaussian noises $\epsilon_{t}$ to the target MR slice $x_{0}^{i+k}$ according to a variance schedule $\beta_{t}$ from $t=0$ to $t=T$, which can be represented as:

\begin{equation}
    q_{t}(x_{t}^{i+k}|x_{t-1}^{i+k})=\mathcal{N}(x_{t}^{i+k};\sqrt{1-\beta_{t}}x_{t-1}^{i+k},\beta_{t}\textbf{I}),
\end{equation}

\noindent Furthermore, we can directly sample $x_{t}^{i+k}$ from $x_{0}^{i+k}$ at an arbitrary timestep $t$ in a closed form using the following accumulated expression:

\begin{equation}
    q_{t}(x_{t}^{i+k}|x_{0}^{i+k})=\mathcal{N}(x_{t}^{i+k};\sqrt{\overline{\alpha}_{t}}x_{0}^{i+k},(1-\overline{\alpha}_{t})\textbf{I}) \Rightarrow x_{t}^{i+k}=\sqrt{\overline{\alpha}_{t}}x_{0}^{i+k}+(1-\overline{\alpha}_{t})\epsilon,
\end{equation}

\noindent where $\overline{\alpha}_{t}={\textstyle \prod_{s=1}^{t}}(1-\beta_{s})$ and $\epsilon\sim \mathcal{N}(\textbf{0},\textbf{I})$.
To gain the generation ability for MR slices, HiFi-Diff learns the reverse diffusion via a parameterized Gaussian process $p_{\theta}(x_{t-1}^{i+k}|x_{t}^{i+k},x_{cond})$ conditioned on the feature pyramid $x_{cond}$:

\begin{equation}
    p_{\theta}(x_{t-1}^{i+k}|x_{t}^{i+k}, x_{cond})=\mathcal{N}(x_{t-1}^{i+k};\mu_{\theta}(x_{t}^{i+k},t,x_{cond}),\sigma_{t}^{2}\textbf{I}),
\end{equation}

\noindent where $\sigma_{t}^{2}$ is a fixed variance and $\mu_{\theta}(x_{t}^{i+k},t,x_{cond})$ is a learned mean defined as:

\begin{equation}
    \mu_{\theta}(x_{t}^{i+k},t,x_{cond})=\frac{1}{\sqrt{1-\beta_{t}}}\left(x_{t}^{i+k}-\frac{\beta_{t}}{\sqrt{1-\overline{\alpha}_{t}}}\epsilon_{\theta}(x_{t}^{i+k},t,x_{cond})\right) 
\end{equation}

\noindent where $\epsilon_{\theta}(x_{t}^{i+k},t,x_{cond})$ denotes the main branch of HiFi-Diff for noise prediction. To generate in-between slices $x_{0}^{i+k}$,  we iteratively compute the denoising process $x_{t-1}^{i+k}=\mu_{\theta}(x_{t}^{i+k},t,x_{cond})+\sigma_{t}z$, where $z\sim \mathcal{N}(\textbf{0},\textbf{I})$.

HiFi-Diff is trained in an end-to-end manner by optimizing the simple variant of the variational lowerbound $\mathcal{L}_{simple}$ with respect to $\theta$ and $\phi$:

\begin{equation}
    \begin{aligned}
    \mathcal{L}_{simple}(\theta, \phi)&=\mathbb{E}_{x_{0}^{i+k},t,x_{cond}}\left[{\left\|\epsilon_{\theta}(x_{t}^{i+k},t,x_{cond})-\epsilon_{t}\right\|}_{2}^{2}\right] \\
    &=\mathbb{E}_{x_{0}^{i+k},t,x_{0}^{i},x_{0}^{i+1},k}\left[{\left\|\epsilon_{\theta}(x_{t}^{i+k},t,\mathcal{F}_{\phi}(x_{0}^{i},x_{0}^{i+1},k))-\epsilon_{t}\right\|}_{2}^{2}\right],
    \end{aligned}
\end{equation}

\noindent where $\mathcal{F}_{\phi}$ parameterizes the proposed HiFE module, $\epsilon_{t}$ is the Gaussian distribution data with $\mathcal{N}(\textbf{0},\textbf{I})$, and $t$ is a timestep uniformly sampled from $[0,T]$. 

\begin{figure}[t]
    \centering
    \includegraphics[width=0.94\textwidth]{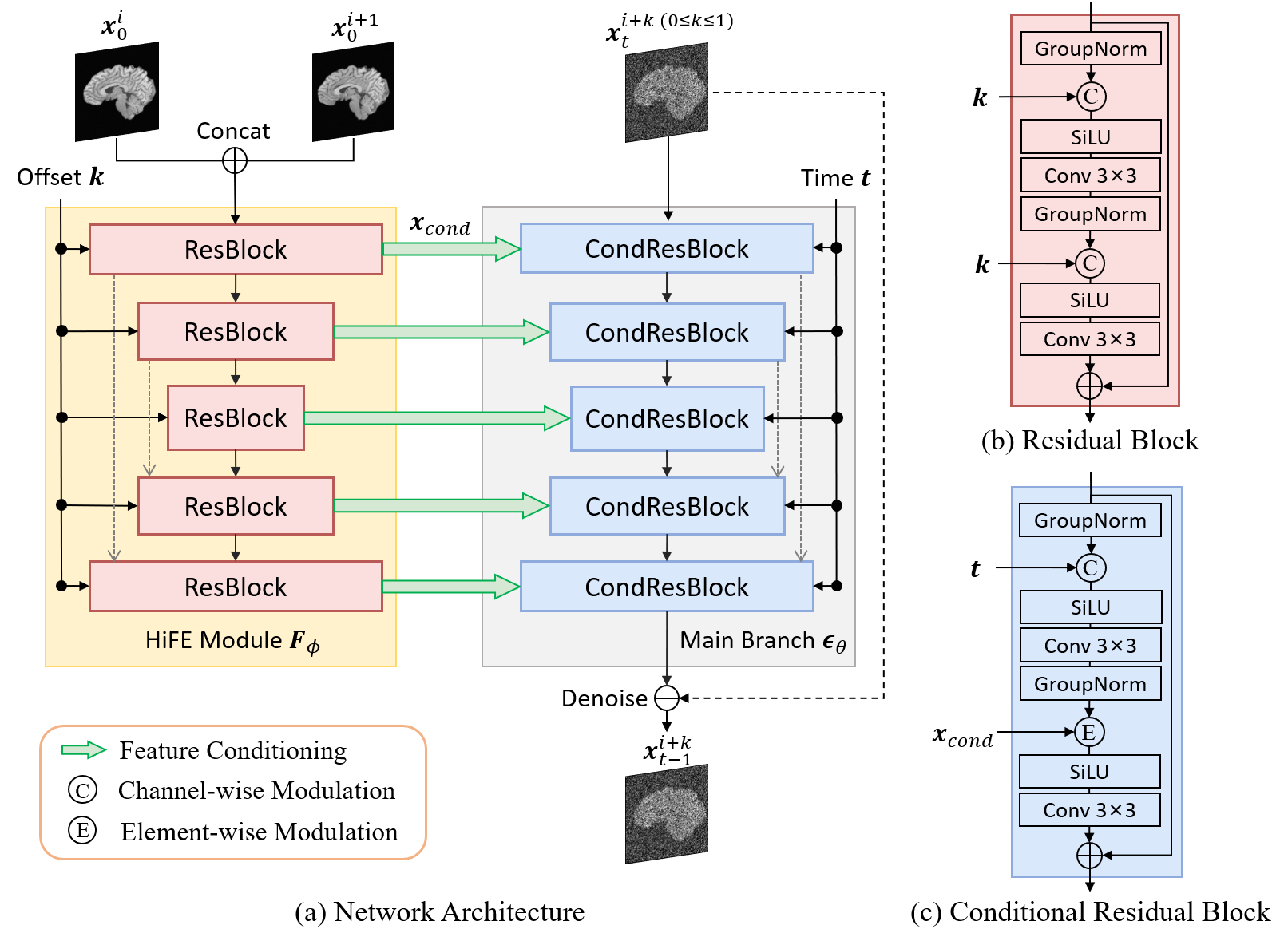}
    \caption{Overview of \textit{\textbf{Hi}erarchical \textbf{F}eature Cond\textbf{i}tional \textbf{Diff}usion (HiFi-Diff)}.}
    \label{fig:overview}
\end{figure}

\subsection{Hierarchical Feature Conditioning Framework}
\label{sec:network}
% Last section briefly introduces the conditional diffusion process (see Fig.~\ref{fig:diffusion}). 
% In this section, we introduce the network architecture of HiFi-Diff, as illustrated in Fig.~\ref{fig:overview}.
Given any pair of $x_{0}^{i}$ and $x_{0}^{i+1}$ from an LR volume with a desired offset $k$, HiFi-Diff is able to iteratively convert a Gaussian noise map into the target in-between slice $x_{0}^{i+k}$ through a reversed diffusion process, as described in the last section.

In this section, we introduce the network architecture of HiFi-Diff. 
As illustrated in Fig.~\ref{fig:overview}(a), the adjacent MR slices $x_{0}^{i}$ and $x_{0}^{i+1}$ are concatenated and input to the proposed HiFE module, which adopts a U-Net~\cite{ronneberger2015u} architecture consisting of a stack of residual blocks (shown in Fig.~\ref{fig:overview}(b)).
% \textcolor{red}{Didn't mention Fig. 3a.}
The offset $k$ is injected into each residual block to perform channel-wise modulation.
Specifically, $k$ is projected by two successive fully connected layers into a 128-dimensional index embedding.
Next, for each layer, the index embedding is passed through a learnable affine transformation to obtain the channel-wise scaling and bias parameters $(k_{s}, k_{b})$, which are applied to the feature map $h$ using the expression ${\rm ChannelMod}(h, k_{s}, k_{b})=k_{s}{\rm GroupNorm}(h)+k_{b}$.
In this way, the HiFE module yields an hourglass-like feature hierarchy that includes feature maps at different scales, with semantics ranging from low to high levels. 

 Conditioned on the timestep $t$ and the feature pyramid $x_{cond}$, the main branch of HiFi-Diff learns to gradually denoise the noise-corrupted input slice $x_{t}^{i+k}$.
The main branch has the same U-Net architecture as HiFE module and consists of a stack of conditional residual blocks.
For each conditional residual block (shown in Fig.~\ref{fig:overview}(c)) in the main branch, the timestep $t$ performs channel-wise modulation in the same way that the offset $k$ does in each residual block of HiFE module.
After the channel-wise modulation, the feature map is further modulated by $x_{cond}$ from the lateral connection at the same image level.
The conditional feature $x_{cond}$ is transformed into the scaling and bias parameters $(x_{s}, x_{b})$ that have the same spatial sizes as the feature map $h$, such that $x_{cond}$ can shift group-normalized $h$ in an element-wise manner: ${\rm ElementMod}(h, x_{s}, x_{b})=x_{s}{\rm GroupNorm}(h)+x_{b}$.
Through element-wise modulation, the hierarchical feature pyramid $x_{cond}$ provides fine-grained conditioning to guide the MR slice generation.

\begin{table*}[t]
\centering
\caption{Quantitative comparison of $\times$4, $\times$5, $\times$6, and $\times$7 SR tasks between the proposed HiFi-Diff and other SR methods.}
\label{tab:compare}
\resizebox{0.95\textwidth}{!}{
\begin{tabular}{c|c|c|c|cc}
\bottomrule%\hline
\multirow{2}{*}{Task} & \multirow{2}{*}{Method}  & \multirow{2}{*}{PSNR} & \multirow{2}{*}{SSIM} & \multicolumn{2}{c}{Dice}                           \\ \cline{5-6} 
                       &                           &                       &                       & WM                                 & GM            \\ \bottomrule %\hline
\multirow{6}{*}{$\times$4}    & \;  Interpolation  \;    & \; $36.05_{\pm2.394}$ \;          & \; $0.9758_{\pm0.0070}$ \;         &  \multicolumn{1}{c|}{ \; $0.9112_{\pm0.0084}$ \;}        & \; $0.7559_{\pm0.0177}$ \;        \\
                       &   DeepResolve             &    \cellcolor{black!25}$39.65_{\pm2.281}$            &   $0.9880_{\pm0.0039}$          &  \multicolumn{1}{c|}{ $0.9700_{\pm0.0021}$  }       &   \cellcolor{black!25}$0.9230_{\pm0.0055}$         \\
                       &   MetaSR                   &   $39.30_{\pm2.287}$             &   $0.9876_{\pm0.0039}$           &  \multicolumn{1}{c|}{ $0.9672_{\pm0.0027}$  }        &   $0.9152_{\pm0.0065}$          \\
                       &   ArSSR                    &   $39.65_{\pm2.282}$             &  $0.9884_{\pm0.0037}$           &  \multicolumn{1}{c|}{  $0.9687_{\pm0.0026}$  }        &   $0.9171_{\pm0.0067}$          \\
                       &   w/o HiFE          &       $38.78_{\pm2.298}$                   &       $0.9874_{\pm0.0039}$           & \multicolumn{1}{c|}{  $0.9639_{\pm0.0033}$  }        &   $0.9118_{\pm0.0078}$        \\
                       &   HiFi-Diff                  &   $39.50_{\pm2.285}$                    &    \cellcolor{black!25}$0.9890_{\pm0.0040}$        &  \multicolumn{1}{c|}{  \cellcolor{black!25}$0.9700_{\pm0.0045}$  }        &   $0.9229_{\pm0.0057}$                 \\ \hline
                       
 \multirow{6}{*}{$\times$5}    & Interpolation   & $31.55_{\pm2.381}$           & $0.9542_{\pm0.0113}$         & \multicolumn{1}{c|}{$0.8362_{\pm0.0090}$} & $0.6470_{\pm0.0136}$ \\
                        & DeepResolve               &  \cellcolor{black!25}$38.28_{\pm2.318}$           & $0.9848_{\pm0.0047}$         & \multicolumn{1}{c|}{$0.9617_{\pm0.0028}$} &  $0.9006_{\pm0.0070}$ \\
                        & MetaSR                    & $37.95_{\pm2.279}$           & $0.9840_{\pm0.0049}$         & \multicolumn{1}{c|}{$0.9522_{\pm0.0043}$} & $0.8808_{\pm0.0100}$ \\
                        & ArSSR                     & $38.27_{\pm2.261}$           & $0.9850_{\pm0.0045}$         & \multicolumn{1}{c|}{$0.9552_{\pm0.0039}$} & $0.8847_{\pm0.0100}$ \\
                        & w/o HiFE&         $37.53_{\pm2.289}$         &          $0.9832_{\pm0.0051}$             & \multicolumn{1}{c|}{$0.9538_{\pm0.0042}$} & $0.8827_{\pm0.0109}$               \\
                        & HiFi-Diff                 & $38.25_{\pm2.260}$           & \cellcolor{black!25}$0.9852_{\pm0.0047}$         & \multicolumn{1}{c|}{\cellcolor{black!25}$0.9620_{\pm0.0060}$} &\cellcolor{black!25}$0.9007_{\pm0.0104}$ \\ \hline
                       
 \multirow{6}{*}{$\times$6}    & Interpolation   & $30.44_{\pm2.379}$           & $0.9457_{\pm0.0129}$         & \multicolumn{1}{c|}{$0.7967_{\pm0.0111}$} & $0.5749_{\pm0.0159}$ \\
                        & DeepResolve               & $37.21_{\pm2.315}$           & $0.9814_{\pm0.0055}$         & \multicolumn{1}{c|}{$0.9521_{\pm0.0033}$} & $0.8786_{\pm0.0078}$ \\
                        & MetaSR                    & $36.55_{\pm2.286}$           & $0.9792_{\pm0.0061}$         & \multicolumn{1}{c|}{$0.9286_{\pm0.0066}$} & $0.8298_{\pm0.0142}$ \\
                        & ArSSR                     & $36.62_{\pm2.271}$           & $0.9798_{\pm0.0058}$         & \multicolumn{1}{c|}{$0.9330_{\pm0.0061}$} & $0.8358_{\pm0.0141}$\\
                        & w/o HiFE &       $36.67_{\pm2.292}$                &                $0.9801_{\pm0.0058}$        & \multicolumn{1}{c|}{$0.9414_{\pm0.0042}$} & $0.8569_{\pm0.0097}$              \\
                        & HiFi-Diff                 & \cellcolor{black!25}$37.41_{\pm2.314}$           &  \cellcolor{black!25}$0.9827_{\pm0.0054}$        & \multicolumn{1}{c|}{ \cellcolor{black!25}$0.9527_{\pm0.0111}$} &  \cellcolor{black!25}$0.8798_{\pm0.0169}$ \\ \hline
                       
 \multirow{6}{*}{$\times$7}    & Interpolation   & $29.61_{\pm2.379}$           & $0.9386_{\pm0.0142}$         & \multicolumn{1}{c|}{$0.7611_{\pm0.0130}$} & $0.5144_{\pm0.0185}$ \\
                        & DeepResolve               & $36.35_{\pm2.303}$           & $0.9782_{\pm0.0063}$         & \multicolumn{1}{c|}{$0.9387_{\pm0.0046}$} & $0.8468_{\pm0.0104}$ \\
                        & MetaSR                    & $35.27_{\pm2.308}$           & $0.9739_{\pm0.0074}$         & \multicolumn{1}{c|}{$0.8988_{\pm0.0092}$} & $0.7691_{\pm0.0180}$ \\
                        & ArSSR                     & $35.20_{\pm2.289}$           & $0.9741_{\pm0.0072}$         & \multicolumn{1}{c|}{$0.9034_{\pm0.0090}$} & $0.7739_{\pm0.0192}$ \\
                        &  w/o HiFE &   $35.86_{\pm2.316}$                     &          $0.9766_{\pm0.0070}$             & \multicolumn{1}{c|}{$0.9245_{\pm0.0056}$} & $0.8254_{\pm0.0122}$              \\
                        & HiFi-Diff                 & \cellcolor{black!25}$36.58_{\pm2.328}$           & \cellcolor{black!25}$0.9797_{\pm0.0062}$         & \multicolumn{1}{c|} { \cellcolor{black!25}$0.9401_{\pm0.0103}$} & \cellcolor{black!25}$0.8550_{\pm0.0177}$ \\ \bottomrule%\hline
                       
\end{tabular}
}
\end{table*}

\section{Experimental Results}
\subsection{Dataset and Experimental Setup}
\subsubsection{Data Preparation}
We collect 1,113 subjects of 3T MR images from the HCP-1200 dataset~\cite{HCP}, with all images having an isotropic voxel spacing of 0.7mm$\times$0.7mm$\times$0.7mm. 
Among these, 891 images are used for training, and the remaining 222 images are used for testing.
We perform N4 bias correction and skull-stripping for preprocessing. It is noteworthy that skull-stripping is necessary in order to protect the privacy of the subjects.
To simulate the LR images with large slice spacing, we downsample the isotropic HR volumes perpendicular to the sagittal view following~\cite{DeepResolve}.

\subsubsection{Implementation Details}
To achieve a comprehensive evaluation, we compare HiFi-Diff with other methods for MR super-resolution, including trilinear interpolation, DeepResolve~\cite{DeepResolve}, MetaSR~\cite{MetaSR}, and ArSSR~\cite{ArSSR}. 
DeepResolve is trained and tested for each specific scaling ratio, while HiFi-Diff, MetaSR, and ArSSR are trained using mixed scaling ratios of ${\times2,\times3,\times4}$ for arbitrary-scale super-resolution.
In each iteration of training, we corrupt the intermediate slice $x_{0}^{i+k}$ with Gaussian noise according to the randomly sampled timestep $t$.
We set $T=1000$ during training, and use DDIM sampler~\cite{DDIM} to speed up the reverse process by reducing $T=1000$ to $T=100$.
All the experiments are conducted using an NVIDIA A100 40G with PyTorch~\cite{pytorch}. 
We use the learning rate of $1.0\times10^{-4}$, batch size of 1, and Adam optimizer~\cite{Adam} to train our model for 700k iterations.

\begin{figure*}[t]
    \begin{center}
    \includegraphics[width=0.92\textwidth]{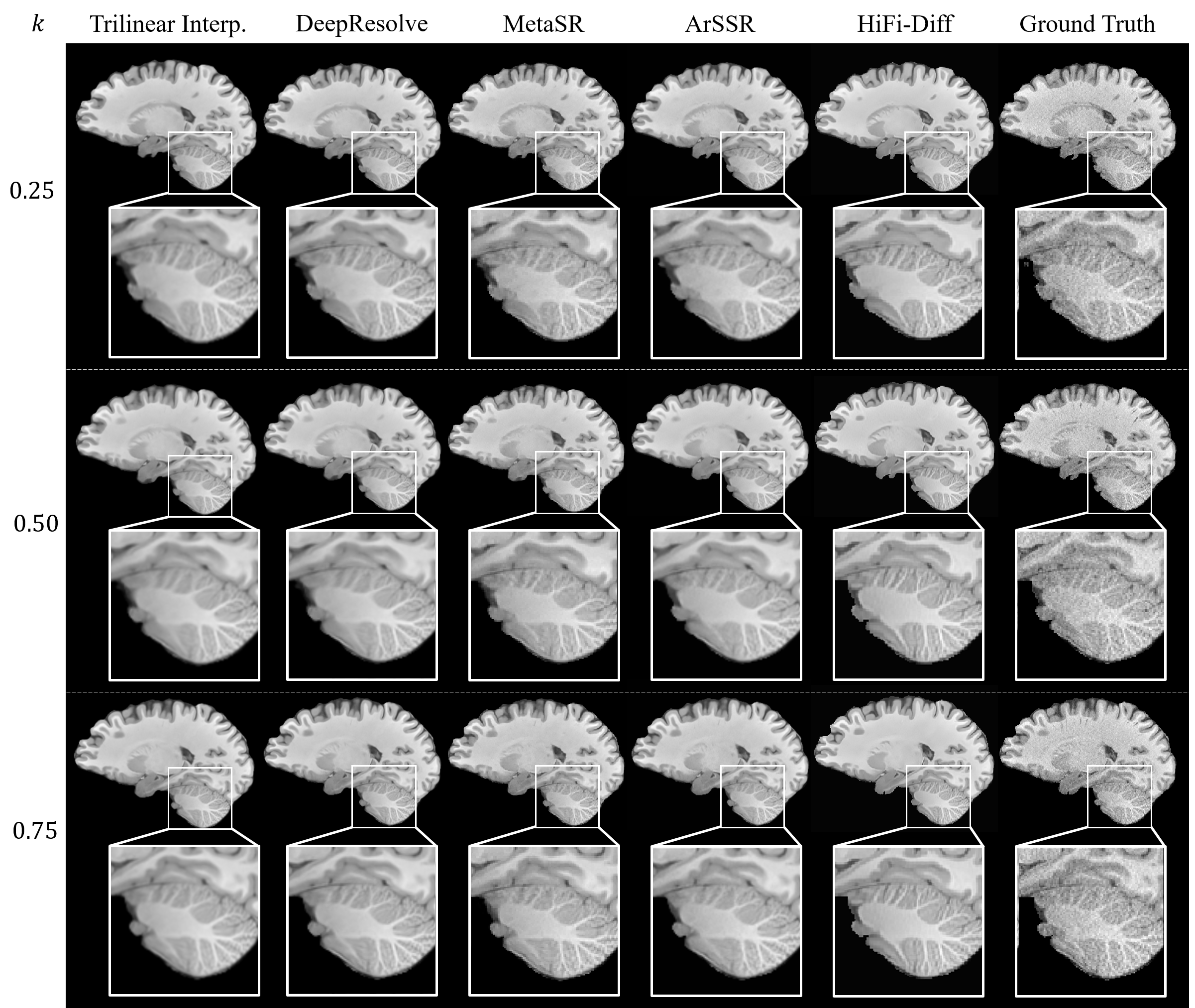}
    \caption{Qualitative comparison of $\times4$ SR task between the proposed HiFi-Diff and other SR methods. The cerebellum regions are highlighted and zoomed in.
    } 
    \label{fig:sr_compare}
    \end{center}
\end{figure*}

\subsection{Super-resolution Evaluation}
We use Peak Signal-to-Noise Ratio (PSNR) and Structural Similarity Index (SSIM) to evaluate the consistency between the SR results and ground truth. Based on the results in Table~\ref{tab:compare}, the proposed HiFi-Diff method outperforms other state-of-the-art methods, particularly at large scaling ratios. It is worth mentioning that DeepResolve achieves the highest PSNR scores at scaling ratios of $\times4$ and $\times5$, which can be attributed to the fact that DeepResolve is specifically trained for each scaling ratio.

In addition, we conduct an ablation study to assess the effectiveness of the HiFE module by removing it and comparing the results.
In detail, we directly inject the concatenated slices into the main branch for element-wise modulation, and concatenate the embedding of offset $k$ and timestep $t$ for channel-wise modulation.
The results show a decrease in all metrics when the HiFE module is removed, indicating that HiFi-Diff benefits from the fine-grained conditioning provided by the HiFE module.

The qualitative comparison of the generated in-between MR slices with different offsets is shown in Fig.~\ref{fig:sr_compare}.
% To qualitatively evaluate the SR performance, we visualize the $\times4$ SR results as shown in Fig.~\ref{fig:sr_compare}.
% We display the sagittal-view slices generated by different SR methods and the ground-truth slices in HR volumes.
% By careful inspection of the cerebellum, trilinear interpolation, DeepResolve, MetaSR, and ArSSR fail to produce complete and clear structures of the white matter.
By inspection of the cerebellum, one can notice that other methods fail to produce complete and clear structures of the white matter for they are optimized using L1/L2 loss, driving their results towards over-smoothing and loss of high-frequency information.
In contrast, HiFi-Diff can faithfully reconstruct image details through an iterative diffusion process.

\begin{figure*}[t]
    \centering
    \includegraphics[width=0.92\textwidth]{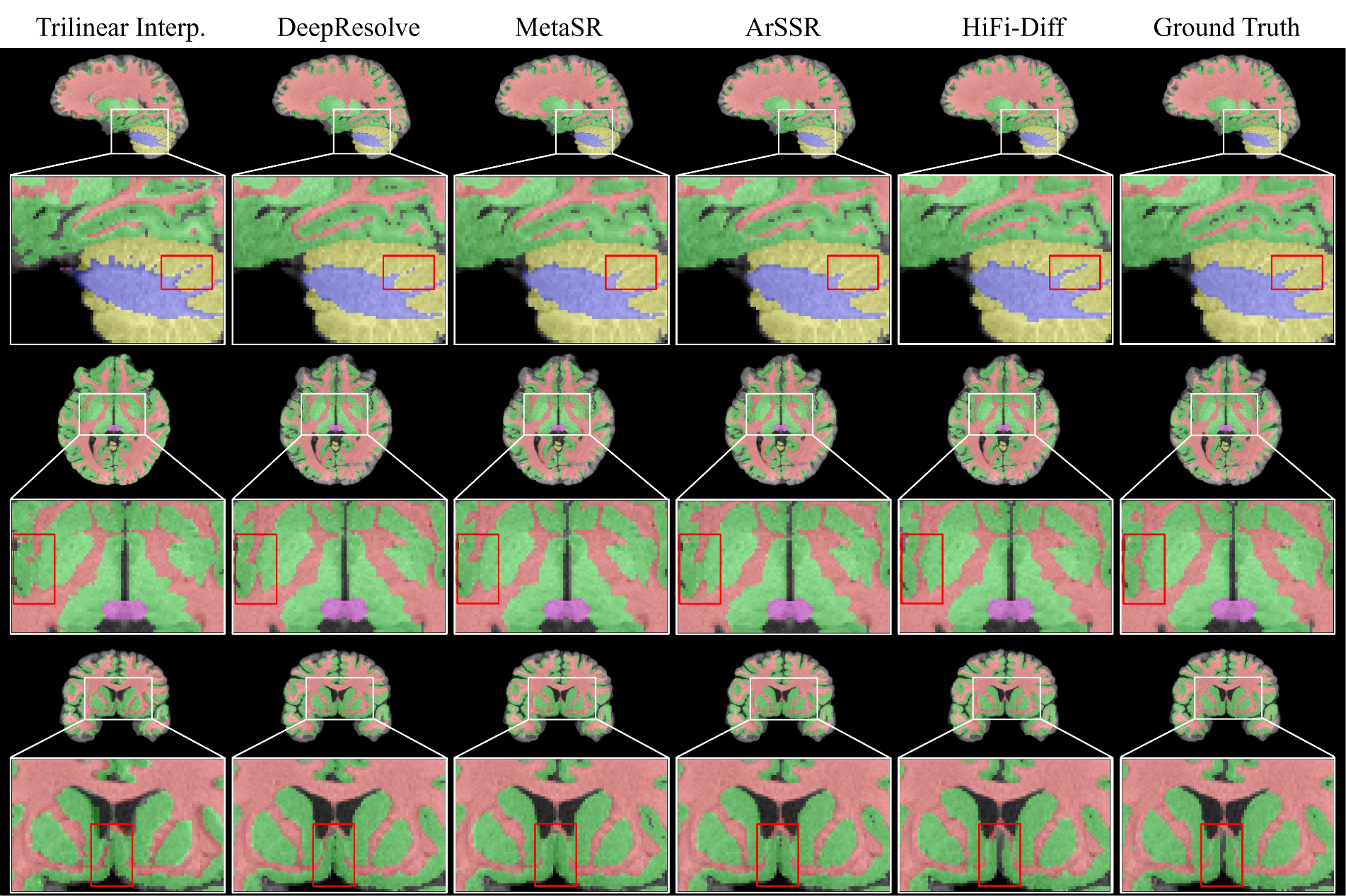}
    \caption{Visual comparison of the fully automatic segmentation on $\times4$ SR results by all the comparing models. The sagittal, axial, and coronal views are shown in three rows, respectively. The areas surrounded by the white boxes are zoomed in below. }
    \label{fig:seg_compare}
\end{figure*}

% \subsection{Evaluation on Downstream Segmentation Task}
To validate the effectiveness of the proposed HiFi-Diff on downstream tasks, we conduct brain segmentation on different SR results using Fastsurfer~\cite{fastsurfer}.
According to Table~\ref{tab:compare}, HiFi-Diff outperforms other methods in terms of Dice score for the white matter (WM) and the gray matter (GM) in most scenarios.
% According to Table~\ref{tab:compare}, HiFi-Diff achieves almost the best Dice scores for the white matter (WM) and the gray matter (GM).
The visual comparison in Fig.~\ref{fig:seg_compare} further demonstrates the superiority of HiFi-Diff, as other methods yield tissue adhesion or discontinuity in their segmented results, while our approach avoids these problems.

\section{Conclusion and Discussion}
In conclusion, we propose HiFi-Diff to conduct arbitrary reduction of MR inter-slice spacing, outperforming previous methods in both generation capability and downstream task performance by leveraging the power of the diffusion models.
To further enhance fine-grained conditioning, we introduce the HiFE module, which hierarchically extracts conditional features and conducts element-wise feature modulations. 
% The relative positional offsets are also provided to handle different scaling ratios of inter-slice spacing.
Despite the superior performance, HiFi-Diff still suffers from slow sampling speed. One possible solution is the implementation of faster sampling algorithms or the utilization of techniques such as knowledge distillation.

% Given two adjacent LR slices, HiFi-Diff is able to generate any in-between MR slices converted from Gaussian noise maps. 
% Furthermore, to handle different scaling ratios of inter-slice spacing, the relative positional offsets are provided to build continuous representations for the spatial positions between two adjacent LR slices. 
% Our experiments on HCP dataset demonstrate that HiFi-Diff is able to generate high-fidelity MR slices that are consistent with the ground truth.
% And the SR results can effectively improve the downstream segmentation performance.

\bibliographystyle{splncs04}
\bibliography{ref.bib}

\end{document}